%% file: SDC-Holography.tex
\documentclass[12pt]{article}
\pdfoutput=1
\usepackage{amsmath,amssymb,amsthm}
\usepackage{bm}
\usepackage[usenames,dvipsnames]{xcolor}
\usepackage{graphicx}
\usepackage{float}
\usepackage{array}
\usepackage{arydshln}
\usepackage{multirow}
\usepackage{rotfloat}
\usepackage{caption}
\usepackage{subcaption}
\usepackage[normalem]{ulem}
\usepackage{cite}
\usepackage{hyperref}
\usepackage{cleveref}

\graphicspath{{figures/}}

\theoremstyle{definition}

\begin{document}

\baselineskip=18pt  
\numberwithin{equation}{section}  
\allowdisplaybreaks  


\vspace*{-2.5cm} 
\begin{flushright}
{\tt IFT-UAM/CSIC-20-142}\\
\end{flushright}

\vspace*{2cm} 
\begin{center}
  {\LARGE\bfseries Tackling the SDC in AdS with CFTs}

 \vspace*{1.5cm}
 {\large Florent Baume$^{1,2}$, Jos\'e Calder\'on Infante$^1$}\\

 \vspace*{1.0cm} 
 { $^1$ Instituto de F\'{\i}sica Te\'orica UAM-CSIC}\par 
 {Cantoblanco, 28049 Madrid, Spain }\par
\vspace{.2cm}
{ $^2$ Department of Physics and Astronomy, University of Pennsylvania}\par
{Philadelphia, PA 19104, U.S.A.}\par
\vspace{.8cm}
\scalebox{.95}{\texttt{fbaume@sas.upenn.edu, j.calderon.infante@csic.es }} \\
  
\end{center}
\vspace*{1.5cm}
%
\noindent 
We study the Swampland Distance Conjecture for supersymmetric theories with
$\text{AdS}_5$ backgrounds and fixed radius through their $\mathcal{N}=2$ SCFT
holographic duals. By the Maldacena--Zhiboedov theorem, around a large class of
infinite-distance points there must exist a tower of exponentially massless
higher-spin fields in the bulk, for which we find bounds on the decay rate in
terms of the conformal data. We discuss the origin of this tower in the gravity
side for type IIB compactification on $S^5$ and its orbifolds, and comment
about more general cases.

\newpage
\tableofcontents

\section{Introduction} \label{sec:intro}

Quantum gravity constraints on effective field theories have recently came
under renewed scrutiny, and several proposals have emerged under the umbrella
of the Swampland Programme \cite{Vafa:2005ui,Ooguri:2006in}, see
\cite{Brennan:2017rbf,Palti:2019pca} for reviews. Among them, the Swampland
Distance Conjecture (SDC) \cite{Ooguri:2006in} is arguably---with the Weak
Gravity Conjecture \cite{ArkaniHamed:2006dz}---one of the most studied facets
of this endeavour. It posits that at points in moduli space located at infinite
geodesic field distance from an arbitrary reference point, there exists a tower
of infinitely many states becoming exponentially light with the geodesic
distance,
\begin{equation}\label{SDCcondition}
	\frac{m}{M_\text{Pl}} \sim e^{-\alpha_{G} \cdot\text{dist}_G}\,,\qquad \text{as dist}_G\rightarrow\infty\, ,
\end{equation}
leading to a breakdown of the effective theory. The distance is computed with
respect to the metric of the moduli space, $G$, and the associated exponential
decay rate, $\alpha_G>0$, is moreover expected to be of order one in Planck
units. For type IIB string theory on a Calabi--Yau three-fold, it has been
possible to put specific lower bounds on $\alpha_G$ by connecting the tower to
BPS states \cite{Gendler:2020dfp}. Similar lower bounds have also been proposed
in conjunction to the Transplanckian Censorship Conjecture
\cite{Bedroya:2019snp,Andriot:2020lea,Bedroya:2020rmd}, and with the RG flow of
BPS strings in 4d $\mathcal{N}=1$ EFTs \cite{Lanza:2020qmt}.

The SDC has been mainly studied in the context of four-
\cite{Palti:2017elp,Heidenreich:2017sim,Blumenhagen:2018nts,Lee:2018urn,Lee:2018spm,Lee:2019tst,Grimm:2018ohb,Grimm:2018cpv}
or six-dimensional \cite{Lee:2019xtm} Minkowski theories with eight or more
supercharges obtained by dimensional reduction of type II string theories, or
their lifts to strong coupling. Using the beautifully-intricate web of
dualities of string theory, it was proposed that the tower of massless states
corresponds to either a decompactification limit or a tensionless weakly-coupled
fundamental string in disguise \cite{Lee:2018urn,Lee:2018spm,Lee:2019tst},
although it may be required to take quantum corrections into account to make
them manifest
\cite{Lee:2019tst,Marchesano:2019ifh,Baume:2019sry,Klaewer:2020lfg}. Note that,
in the latter case, the tower of states generically contains arbitrarily large
higher-spin fields. See \cite{Scalisi:2019gfv} for implications in (quasi-)dS
spaces.

A variation of this framework is the inclusion of a potential
\cite{Baume:2016psm,Klaewer:2016kiy,Valenzuela:2016yny,Blumenhagen:2017cxt,Hebecker:2017lxm,Buratti:2018xjt,Landete:2018kqf,Ooguri:2018wrx,Hebecker:2018fln,Gonzalo:2018guu,Lee:2019tst,Blumenhagen:2019qcg,Font:2019cxq,Grimm:2019ixq},
which may lift the flat spacetime geometry to an AdS space. In the limit of
large AdS radius in Planck units, $LM_\text{Pl}\to\infty$, a similar behaviour
is expected, with an infinite tower of states also becoming massless, behaving
as $m/M_\text{Pl}\sim (LM_\text{Pl})^{-\alpha}\,, \alpha>0$
\cite{Lust:2019zwm}.  In supersymmetric cases a strong version of the
conjecture suggests $\alpha = \frac{1}{2}$, usually interpreted as a
consequence of the no-scale-separation condition between the internal manifold
and the AdS radius. In string-theoretic realisations of these AdS geometries,
the tower is often identified with a sector of Kaluza--Klein modes. Part of the
internal manifold and the AdS space are stabilised by the same fluxes and, as a
consequence, the AdS radius and a breathing mode of the compact space are
linked together. The limit of large radius will then also lead to a
decompactification. For recent works, see
\cite{Font:2019uva,Junghans:2020acz,Buratti:2020kda,Marchesano:2020qvg,Lust:2020npd}.

This proposal is somewhat different from what one would naively call the
Swampland Distance Conjecture for moduli spaces of AdS vacua. Even though it is
exploring the possible AdS vacua of the theory, it is not about the
continuously-connected part parametrised by massless scalars, which we refer to
in this work as the moduli space. It is rather about the different branches of
vacua parametrised by massive scalars. In string theory constructions, the
presence of fluxes will give masses to the scalars controlling these limits and
can therefore no longer be considered as moduli in the usual sense. Typically
one consider different branches of vacua in this setup by changing the flux
quanta.

This raises the question of whether the SDC extends to moduli spaces of AdS
vacua in the sense described above, and what kind of towers of states can be
expected to appear. In those setups, the AdS scale in Planck units,
$LM_\text{Pl}$, remains fixed throughout all the moduli space. This is the kind
of trajectories we want to tackle in this work.

In this context, an open question is whether it is possible to consider
decompactification limits. Such trajectories would imply the possibility of
tuning the size of an internal dimension without changing the AdS radius at
all.  Current models featuring a separation of scales always link the AdS
radius and the internal dimensions in some way, while the limits we are
interested in would require them to be independent. Although inconclusive,
current understanding of AdS vacua seems to disfavour such trajectories and
leads to the intriguing possibility that equi-dimensional and
non-equi-dimensional limits in AdS are distinguished, being called SDC and ADC
directions, respectively.

In the case of equi-dimensional limits it is reasonable to expect the
appearance of tensionless strings. However, an immediate challenge one faces is
that those points are out of the regime of parametric control of the usual
supergravity description of these vacua. Indeed, the tension of these strings
will eventually fall below the AdS scale, leading to the notoriously difficult
problem of quantising strings in highly-curved backgrounds. To retain control
over the theory one conversely assumes a weakly-curved background,
corresponding to a semi-classical approximation. A possible way to go around
the issue is to make use of the AdS/CFT correspondence \cite{Maldacena:1997re}.
Our aim here is to analyse a possible extension of the SDC by studying the
evolution of physical quantities through their CFT duals.

Using holography as a tool to study the Swampland programme has already bore
fruits. Proofs of no-go theorems applied to global symmetries as well as the
Weak Gravity Conjectures in AdS were established by relating black hole
quantities to conformal data \cite{Harlow:2018tng,
Harlow:2018jwu,Montero:2018fns}. More recently, positivity bounds were related
to moduli stabilisation constraints in AdS${_4}$/CFT${}_3$, as well as possible
connections to the SDC \cite{Conlon:2020wmc}. Closer to our setup, the classical
moduli space has further been shown to be a coset in the case of $\text{AdS}_5$
gauged supergravity with sixteen (real) supercharges \cite{Louis:2015dca}.

\subsection*{Moduli and Marginal Deformations}

In AdS/CFT, each field of mass $m$ in the bulk is associated on the boundary to
a conformal operator of dimension $\Delta$. The dictionary between scalars and
$\ell$-symmetric traceless tensors is given in AdS$_5$/CFT$_4$ by:
\begin{align}
	m^2L^2 =&\,\Delta(\Delta-4)\,, & \text{(scalars)}\,; \label{dicScalar}\\
	m^2L^2 = &\,(\Delta +\ell - 2)(\Delta - \ell - 2)\,,  & (\ell\text{-symmetric traceless tensors})\label{dic-pform}.
\end{align}
The variation of any mass as a function of the moduli in the bulk can thus be
controlled by tuning parameters of the CFT. We note that, as we demand the AdS
radius, $L$, to be fixed in Planck units, we can use these expressions to
evaluate the mass of a given state in Planck units up to some numerical
coefficients which are irrelevant to our analysis.

The moduli space, parametrised by the vacuum expectation value (vev) of the
moduli, $z^i$, is then identified with the conformal manifold, the space of
exactly marginal deformations, $\lambda^i$, of the CFT, see e.g.
\cite{Tachikawa:2005tq,Hertog:2017owm}. This conformal manifold is endowed with
the so-called Zamolodchikov metric, $\chi$, that is the dual of the bulk moduli
space metric, $G$:
\begin{equation} \label{eq:duality-metrics}
	\left(\mathcal{M}_{\text{mod}}, G_{ij}(z)\right) \longleftrightarrow \left(\mathcal{M}_\text{CFT}, \chi_{ij}(\lambda)\right)\,.
\end{equation}

More specifically, the Zamolodchikov metric is mapped to the metric in moduli
space measured in AdS units up to a constant prefactor, such that in Planck
units we have:
\begin{equation} \label{eq:metric-dictionary}
	(LM_\text{Pl})^3\, G_{ij}(z) \sim  \chi_{ij}(\lambda)\, .
\end{equation}
The specific form of the metric can be computed as a series in large $N$, dual
to the weakly-coupled quantum-gravity expansion in the bulk
\cite{Nojiri:1998dh}. As the partition function of the bulk on the boundary is
identified with the generating functional of correlation functions of the CFT,
corrections on either side are guaranteed to match on the other. This means
that we can use \eqref{eq:metric-dictionary} to compute, at least in principle,
the metric in moduli space from the Zamolodchikov metric in any regime of the
theory.

As it is well known, unitarity furthermore imposes constraints on the CFT data.
For instance the dimension is bounded from below, and an operator with spin
$\ell$ must satisfy:
\begin{equation}\label{unitaritybound}
	\begin{aligned}
		\Delta& \geq 1\,,\qquad\qquad \ell=0\,;\\
		\Delta& \geq \ell+2\,,\qquad\, \ell>0\,.
	\end{aligned}
\end{equation}
For $\ell=0$ the bound is saturated by a free scalar field and maps to
tachyonic fields in the bulk, while the flavour currents and the
energy-momentum tensor sit at the bound for $\ell=1\,,2$, respectively. For
$\ell>2$, the bound is associated to so-called higher-spin conserved currents.
By the Maldacena--Zhiboedov theorem \cite{Maldacena:2011jn} and its extensions
\cite{Stanev:2013qra,Boulanger:2013zza,Alba:2013yda,Li:2015itl,Hartman:2015lfa},
having a single higher-spin conserved current implies the existence of an
infinite number of them. Moreover this can only occur in the presence of
(generalised) free fields. Such higher-spin currents will be a central part of
this work.

With this framework, the study of the Swampland Distance Conjecture for AdS
spacetime can then be rephrased as an analysis of the possible infinite-distance
points of conformal manifolds. We will focus on four-dimensional $\mathcal{N}=2$
SCFTs, corresponding to supergravity theories with sixteen (real) supercharges
on $\text{AdS}_5$, although we will comment on implications in other dimensions.
In those theories, $(\mathcal{N}=2)$-preserving exactly marginal deformations
can only correspond to variations of (complexified) gauge couplings \cite{Argyres:2015ffa, Cordova:2016xhm}, and it is
then easy to give a physical interpretation to the towers of states in terms of
gauge data.

In particular, we will be able to track how the dimension of certain operators
behaves near a class of infinite-distance points corresponding to
weakly-coupled gauge subsectors of the SCFT. Among them, we find the
above-mentioned infinite tower of higher-spin currents that saturate the
unitarity bound \eqref{dic-pform} in the presence of free fields. In the bulk,
this leads to an infinite tower of higher-spin fields becoming massless and
satisfying the condition \eqref{SDCcondition}. We will further estimate the
decay rate, that will be shown to be \emph{at least} of order one.

This work is structured as follows: in section \ref{sec:warm-up} we describe how
to use AdS/CFT to study the SDC in the bulk by working in the moduli space of
AdS$_{5}\times$S$^{5}$ vacua of type IIB string theory. The high degree of
supersymmetry is enough to compute the metrics exactly and therefore
constitutes the simplest example to study. In section \ref{sec:CFTs} we review
some important properties of the conformal manifolds associated to
four-dimensional $\mathcal{N}=2$ SCFTs and study the SDC on general grounds. In
section \ref{sec:orbifolds} we make contact again with the bulk by examining a
specific family of $\mathcal{N}=2$ SCFTs with known bulk duals. We give our
conclusions and discuss applications to other dimensions in section
\ref{sec:conclusions}. Additionally, we briefly review unitary representations
of the $\mathcal{N}=2$ superconformal algebra in appendix
\ref{app:representations}.

\paragraph{Note added:} When finalising the preparation of this article, we
became aware of an upcoming work by Perlmutter, Rastelli, Vafa, and 
Valenzuela \cite{Perlmutter:2020buo}, which explores similar ideas.

\section{A Warm-up: Type IIB $\text{AdS}_5\times \text{S}^5$ Vacua} 
\label{sec:warm-up}

As a first example and to set our nomenclature and conventions, we consider the
family of $\text{AdS}_5$ vacua obtained by compactifying Type IIB on $S^{5}$
with $N$ units of $F_{5}$-flux. It is the most celebrated example of the
holographic principle, being dual to $\mathcal{N}=4$ super-Yang--Mills theory
with gauge group $SU(N)$ in four dimensions. Due to the high amount of
supersymmetry, non-renormalisation theorems makes it possible to compute
relevant quantities exactly. 

This case will be useful to exemplify the inclusion of the moduli space into
the scalars manifold, including the stabilised scalar fields. It will
make a clear distinction between the limits we want to explore and the ``ADC
directions'', where the AdS scale, $L$, is allowed to vary \cite{Lust:2019zwm}.
It will also serve to illustrate the issues of parametric control that occur
when approaching infinite-distance points in AdS moduli space and how the dual
CFT picture allows one to circumvent them.

This family of solutions is parametrised by $N$ and the complex axio-dilaton,
\begin{equation}
	\tau = C_{0} + i\frac{1}{g_s}\,,
\end{equation}
made out of the string coupling, $g_s$, and the Type IIB axion, $C_0$. The AdS
radius, $L$, is forced to coincide with the radius of the five-sphere and is set
to:
\begin{equation} \label{eq:AdS-scale-N=4}
	L^4 = 4\pi g_s N {\alpha'}^2 = \frac{1}{4\pi^3} g_s N M_{s}^{-4}\, .
\end{equation}
In terms of the five-dimensional Planck mass, it is rewritten as
\begin{equation}\label{eq:AdS/Plank-scale}
	L\,M_\text{Pl} \sim N^{2/3}\,,\qquad
	M_\text{Pl} \sim g_{s}^{-1/4} N^{5/12} M_{s}\,,
\end{equation}
so that keeping the AdS scale fixed in Planck units corresponds to
fixing $N$. Thus, the moduli space of AdS vacua is parametrised solely by the
axio-dilaton.

From the perspective of compactification, this is understood as a stabilisation
of the scalar associated to the breathing mode of the sphere through fluxes,
such that it is no longer a modulus. Note that with the nomenclature
established in the introduction, this stabilised scalar is associated to the
ADC direction and not part of the moduli space of massless scalars.

Computing the moduli space metric usually requires one to perform the
dimensional reduction of type IIB supergravity on $S^5$, including kinetic
terms for the axio-dilaton and the breathing mode, and substitute
\eqref{eq:AdS-scale-N=4}.\footnote{According to the generalised SDC
\cite{Lust:2019zwm}, this should also include the contribution to the distance
due to the change of the AdS scale, which is irrelevant here as it is kept
fixed.} However, we can here take full advantage of type IIB S-duality which
is preserved in this background, and constrains the metric to be:
\begin{equation} \label{axio-dilaton metric}
	ds^2 \sim \frac{d\tau d\bar{\tau}}{\text{Im}(\tau)^2}\,,
\end{equation}
up to numerical factors that will be irrelevant for our purposes. It is then
obvious that there are only two infinite-distance points: $\text{Im}\tau\to
0,\infty$, which are physically equivalent. We focus the rest of
the discussion on the latter.

One could naively expect the SDC to work exactly as it does in flat space: using
the metric \eqref{axio-dilaton metric} any geodesic approaching $\text{Im}\tau
\to \infty$ is forced to move only along the $\text{Im}\tau$-direction, and as a
consequence the distance behaves logarithmically with $g_s$. The associated
tower of states is then identified with string excitations controlled by the
string scale, $M_s$, which falls polynomially to zero in Planck units. Putting
the two together, we find the expected exponential behaviour.

However, for this argument to work a key point is to remain under parametric
control along the trajectory. In particular the quantisation of the string with
the usual methods requires one to be in the \emph{weakly-curved} regime. This
imposes the string scale to be above the AdS scale:
\begin{equation}
	LM_s \sim \left( g_{s} N\right)^{1/4} \gg 1\,.
\end{equation}
As the AdS scale is fixed along the trajectory, we are no longer under
parametric control as $\text{Im}\tau\to \infty$. This extra condition does not
arise when considering the moduli spaces of Minkowski vacua such as those
considered in the usual compactification to flat backgrounds.

We are therefore leaving the phase of the moduli space where the supergravity
description is valid, making a qualitative assessement of the SDC impossible,
as the two infinite-distance points, $\tau = 0,i\infty$, are both inaccessible
in that regime. However, we are conversely entering a phase where a weakly-coupled
description in terms of the conformal theory is appropriate. There, the bulk
axio-dilaton is identified with the complexified gauge coupling,
\begin{equation}
	\tau = \tau_\text{YM}= \frac{\theta}{2\pi}+i\frac{4\pi}{g^{2}_\text{YM}}\,,
\end{equation}
and parametrises the only possible $(\mathcal{N}=4)$-preserving marginal
deformation. Due to the amount of supersymmetry, the Zamolodchikov metric,
$\chi$, is found to be quantum exact and can be computed through usual
diagrammatic methods, or by localisation techniques reviewed next section:
\begin{equation}\label{N=4Zamolochikov}
	\chi_{\tau\bar{\tau}} \sim \frac{N^2-1}{\text{Im}(\tau)^2}\,.
\end{equation}
The numerator is set by the dimension of the gauge group $\mathcal{G}=SU(N)$,
and we once again ignored irrelevant order one prefactors.  As expected from
the bulk, there are also two physically-equivalent infinite-distance points
related by S-duality, and the bulk limit $\text{Im}\tau\to \infty$ corresponds
to a free theory, $g_\text{YM}=0$, on the CFT side.

The operators of the CFT are gauge-invariant composite operators made out of
fields in the $\mathcal{N}=4$ vector multiplet.\footnote{For simplicity, we do
not take into consideration the R-symmetry structure of these fields,
and generically denote any of the scalars transforming in the
$\mathbf{6}$ of $SU(4)_R$, or later any other scalar, by $\phi$. For our
purpose, it will be irrelevant.} Their conformal dimensions are given by the
sum of the free value and their anomalous dimension, $\gamma$:
\begin{equation}
	\Delta = \Delta_\text{free} + \gamma(\tau)\,.
\end{equation}
In the free limit, the conformal dimension is obtained by naive dimensional
analysis. For instance, the lowest-lying operators is given by
$\text{Tr}\phi^2$ and is of dimension $\Delta=2$ in the limit
$\text{Im}\tau\to\infty$. Using the dictionary \eqref{dicScalar}, it therefore
corresponds to a field at the Breitenlohner-Freedman (BF) bound in the bulk.

From the SDC one expects a tower of states becoming massless exponentially with
the distance at the infinite-distance point. To see what happens on the CFT
side, we can use perturbation theory to write the leading contribution in the
$g_{YM}\to 0$ limit as
\begin{equation} \label{eq:exponential-CFT}
	\Delta = \Delta_\text{free} + \eta\, g_{\text{YM}}^{\beta} + 
	\mathcal{O}(g_{\text{YM}}^{\beta+1}) =
	\Delta_\text{free} + \eta\, e^{-\alpha_{\chi}\, \text{dist}_\chi(\tau)}\, ,
\end{equation}
where $\alpha_\chi\,,\beta\,,\eta$ are coefficients depending on the type of
operator considered. In the second equality we have used the expression for the
distance with respect to the Zamolodchikov metric in terms of the Yang--Mills
coupling. We can easily see that---with the exception of operators whose
dimensions are protected by a selection rule---the conformal dimension falls
exponentially fast to its free value. An important class of such operators are
spin-$\ell$ operators of the form:
\begin{equation}\label{higherSpinCurrents}
	J_{\mu_1\dots\mu_{\ell}} = 
	\bar{\phi}\;\overleftrightarrow{\partial}_{(\mu_1}\cdots
	\overleftrightarrow
	{\partial}_{\mu_\ell)}\phi - \text{(traces)}\,.
\end{equation}
These operators have an anomalous dimension at a generic point of the conformal
manifold but---using the equations of motion---become conserved in the free
limit and saturates the unitarity bound \eqref{unitaritybound}. The presence of
these higher-spin conserved currents in a CFT in fact implies that the theory is
free by the Maldacena--Zhiboedov theorem \cite{Maldacena:2011jn}.

In the bulk they are identified with higher-spin fields that become massless
exponentially fast:
\begin{equation} \label{higher-spin-fields}
	M^2_\ell L^2 = (\Delta+\ell-2)(\Delta-(\ell+2))\sim e^{-\alpha_{G}\, 
	\text{dist}_G(\tau)}\,.
\end{equation}
We therefore indeed have a tower of massless modes in the bulk when going to
the infinite-distance point, behaving according to the Swampland Distance
Conjecture. 

However, the unitarity bound \eqref{unitaritybound}
implies that fields dual to scalar operators of the CFT---e.g.  single-trace
operators, $\mathcal{O}\sim \text{Tr}\phi^n$---remain massive in the bulk and
are regularly spaced for sufficiently large $n$:
\begin{equation}\label{singleTraceDecay}
	M^2_\text{scal.} L^2 \sim n^2+\mathcal{O}(e^{-\text{dist}(\tau)})\,.
\end{equation}
This is a striking difference with respect to the usual results of the SDC for
Minkowski backgrounds: in this case the tower is formed by higher-spin modes,
which are in principle interacting,\footnote{For a review of the higher-spin/CFT
duality, see e.g. \cite{Giombi:2016ejx}} but there are only a small number of
massless scalar fields. In flat space, the tower always contains an infinite
number of massless scalars. These residual masses in our setup are likely
related to the presence of curvature and fluxes.

The origin of the tower is however clear: as $g_s\to0$, the higher-spin fields
are those expected from a tensionless fundamental string. The density of the
tower is moreover linear, $M_{\ell}^{2}\sim \ell$, which agrees with the flat
space expectation, while that of a Kaluza--Klein tower is $M_{k}^{2}\sim k^{2}$
for sufficiently large $k$ \cite{Lee:2019wij}. As we elaborate in the following
section, this is a very generic behaviour when a tower of higher-spin conserved
currents appears in the CFT, and lends credence to the expectation that
infinite-distance points at fixed AdS radius should not be decompactification
limits.

One can also ask about the order of magnitude of $\alpha_{G}$ in equation
\eqref{higher-spin-fields}. From \eqref{N=4Zamolochikov} we see that
$\alpha_{\chi}$ in equation \eqref{higherSpinCurrents} is, up to order one
factors, given by $\alpha_{\chi} \sim \text{dim}(SU(N))^{-1/2}$. However, we
recall that the relation between the moduli space metric and that of the
conformal manifold \eqref{eq:metric-dictionary} introduces a dependence on
$(LM_\text{Pl})$, which in turn depends on $N$ \eqref{eq:AdS/Plank-scale}. This
factor enters in the relation between $\alpha_{\chi}$ and $\alpha_{G}$, which
is nothing but taking into account that they are measured in AdS and Planck
units. All in all, we find that the exponential rate is order one
in Planck units:
\begin{equation} \label{eq:relation-alphas}
	\alpha_{G} \sim (LM_\text{Pl})^{3/2}\, \alpha_{\chi} \sim \mathcal{O}(1) \, .
\end{equation}

Before closing this section, let us come back to the issue of a well-defined
supergravity description. Parametric control is lost when $LM_s<1$, the scale
at which the tower of higher-spin modes falls below the one set by the AdS
radius.  To obtain an effective description in that regime, it would be
required to integrate out these fields in a consistent way, and the new cut-off
of the theory would be below the AdS scale. This does not seem to be a
meaningful description of the physics in AdS. This suggests that an infrared
description of quantum gravity in terms of $\text{AdS}_5$ supergravity is not
appropriate to describe an infinite-distance limit, and such a point has to be
located at the boundary of the quantum moduli space. One is instead forced to
go to the CFT dual to probe such a limit, where the theory is free.

While $\mathcal{N}=4$ super-Yang--Mills and type IIB $\text{AdS}_5\times S^5$
vacua are very constrained by symmetries and their respective metrics can be
understood throughout the entire moduli space, they illustrate a behaviour that
is quite universal: when a subsector of the theory becomes free, an infinite
number of higher-spin conserved currents always appear at that point in the
conformal manifold. In addition, single-trace operators, as the ones we used in
equation \eqref{singleTraceDecay}, are omnipresent in conformal gauge theories.
It is also general that there can only be a small number of scalar fields
sitting at the BF bound, as the dual operators must take the form
$\text{Tr}(\phi^2)$.

\section{$\mathcal{N}=2$ Conformal Manifolds in Four Dimensions} \label{sec:CFTs}

Strengthened by the observations made in the previous section, we would now
like to extend these arguments to theories with less supersymmetry. We will
focus on theories with sixteen real supercharges in the bulk, in
particular those obtained by compactifying on an orbifold of $S^5$. In the SCFT
dual, half of the five-dimensional supercharges are mapped to superconformal
generators, and one obtains four-dimensional $\mathcal{N}=2$ SCFTs. Before
studying the infinite-distance points in both description in more details, let
us review some well-established facts about $\mathcal{N}=2$ theories.

As mentioned above, to be able to define a notion of distance on the conformal
manifold, $\mathcal{M}_{CFT}$, the relevant object is the so-called
Zamolodchikov metric, $\chi$. Denoting the set of all exactly marginal
operators by $\mathcal{O}_i$ and their associated coupling
constants by $\tau_i$, it is defined as the coefficient of the two-point
correlators of marginal operators:
\begin{equation}\label{ZamolodchikovMetric}
	\left<\mathcal{O}_i(x)\mathcal{O}_{\bar{\jmath}}^\dagger(y)\right> = 
	\frac{\chi_{i\bar{\jmath}}(\tau)}{|x-y|^{8}}\,.
\end{equation}

Supersymmetry as well as the number of spacetime dimensions constrain the
structure of the superconformal multiplets, the possible marginal deformations,
and the properties of the metric. For $\mathcal{N}=2$, a relevant class of
multiplets are the chiral (resp. anti-chiral) multiplets, denoted
$\mathcal{E}_r$ (resp. $\bar{\mathcal{E}}_{-r}$), with $r$ their $U(1)_R$
charge.  They have the property of being annihilated by four of the
supercharges:\footnote{We use the nomenclature of $\mathcal{N}=2$
superconformal multiplets of \cite{Dolan:2002zh } and, by abuse of notation,
also denote their superconformal primaries by $\mathcal{E}_r$.}
\begin{equation}
	[\bar{Q}_{\dot{\alpha}a},\mathcal{E}_r]=0\,,\qquad
	\Delta=r\,.
\end{equation}
Our convention for the quantum numbers and the relevant notions pertaining to
superconformal multiplets and their primaries are reviewed briefly in appendix
\ref{app:representations}.

These multiplets form a ring under the operator product expansion, and can be
used to probe a host of properties of a given SCFT. For us, their importance
comes from the fact that the chiral ring contains the only possible marginal
operator preserving eight supercharges.

To be able to define the $\mathcal{N}=2$ Zamolodchikov metric
\eqref{ZamolodchikovMetric}, an operator, $\mathcal{O}$, must satisfy the
following properties: be exactly marginal, $\Delta_\mathcal{O}=4$; be a singlet
under the R-symmetry group, $SU(2)\times U(1)$, namely $(R, r)= (0, 0)$; and
annihilated by all supercharges, $Q\,,\bar{Q}$ (up to total derivatives).  Note
that such an operator need not be a \emph{super}conformal primary, but simply a
conformal primary. It turns out that in the case of four-dimensional
$\mathcal{N}=2$ theories the only such operator is the bottom component of
$\mathcal{E}_2$ (or its conjugate) \cite{Argyres:2015ffa, Cordova:2016xhm}.
This operator is indeed by definition annihilated by all anti-chiral
supercharges, $\bar{Q}$, and is reached from the superconformal primary by
successive applications of the four remaining supercharges, $Q$. As each
application of a supercharge increases the conformal dimension by $1/2$, it is
also exactly marginal, $\Delta=4$.  One can further verify that all these
marginal operators are then proportional to $\theta$- or gauge kinetic terms,
\begin{equation}
	\mathcal{O} = Q^4\mathcal{E}_{2}\sim \text{Tr}(F\wedge\ast F+i 
	F\wedge F)\,.
\end{equation}
When a Lagrangian description is available, the deformation term is obtained by
integrating the multiplet over superspace and corresponds to an F-term:
\begin{equation}
	\delta\mathcal{L}=\tau^i \int d^4\theta(\mathcal{E}_{2})_{i} +c.c.
\end{equation}
As such, the only possible marginal deformations preserving $\mathcal{N}=2$
correspond to a modification of Yang--Mills couplings. 

\subsection{The Zamolodchikov Metric} \label{sec:The-Zamo-Metric}

From the discussion above, the conformal manifold metric is therefore related
to the two-point functions of the superconformal primaries of the chiral
multiplets, $\left<\mathcal{E}_{2,i}\,\bar{\mathcal{E}}_{-2,j}\right>$. The
structure of the conformal manifold of four-dimensional $\mathcal{N}=2$ SCFTs
is extremely constrained. It was indeed shown that superconformal symmetry
imposes the conformal manifold to be Hodge--K\"ahler, and that its K\"ahler
potential, $K$, is related to the partition function on the
four-sphere\cite{Gomis:2014woa, Gerchkovitz:2014gta,Gomis:2015yaa}:
\begin{equation} \label{eq: Zamo-partition-function}
	\chi_{i\bar{\jmath}}= 192\,\partial_i\bar{\partial}_{\bar{\jmath}} K\,,\qquad
	K= 12\log(Z_{S^4})\,.
\end{equation}
This is a very powerful statement, as the four-sphere partition function of
such theories can then be computed via localisation techniques
\cite{Pestun:2007rz}. Indeed, if the SCFT has a Lagrangian description anywhere
in the conformal manifold, the partition function can be written as a an
integral over the Cartan subalgebra, $\mathfrak{h}$, of the gauge group:
\begin{equation}
	Z_{S^4}(\tau_i,\bar{\tau}_i) =  \int_\mathfrak{h}da \,
	\Delta(a)\left|Z_\Omega(a,\tau_i)\right|^2\,,
\end{equation}
where $\Delta(a)$ is the Vandermonde determinant, and the integrand factorises
as
\begin{equation}\label{PartitionFunctionContribution}
	Z_\Omega = Z_{\Omega,\text{cl}}(a,\tau_i) \cdot 
	Z_{\Omega,\text{loop}}(a)\cdot Z_{\Omega,\text{inst}}(a,\tau_i)\,.
\end{equation}
The classical contribution is universal, 
\begin{equation}\label{Zcl}
	\left|Z_{\Omega,\text{cl}}(a)\right|^2 = \exp\left( 2\pi 
	\text{Im}(\tau)\text{Tr}a^2 \right)\,,
\end{equation}
while the one-loop and instanton contributions depend on the spectrum of the
theory under consideration. For the special subset of $\mathcal{N}=4$
super-Yang--Mills theories, there are no one-loop or instanton contributions,
and the computation is reduced to performing a Gaussian integral:
\begin{equation} \label{N=4-part-function}
	Z^{\mathcal{N}=4}_{S^4}(\tau,\bar{\tau}) \sim (\text{Im}\tau)^{-\text{dim}(\mathcal{G})/2}\,.
\end{equation}
Taking derivatives, one arrives again to the result advertised in equation
\eqref{N=4Zamolochikov}.

For a generic $\mathcal{N}=2$ theory one can obtain the metric as a formal
power series by performing an expansion with respect to marginal couplings.
This technique has been used to find the perturbative expansion of the
Zamolodchikov metric to high order in SQCD \cite{Baggio:2014ioa,
Gerchkovitz:2014gta} and the large-$N$ limit of necklace theories
\cite{Pini:2017ouj}.

\subsection{The SDC and Weakly-gauged Points}

It is easy to see that at any point of the manifold for which a subset
$\{\tau_a\}$ of the marginal couplings go to the free limit,
$\text{Im}\tau_a\to\infty$, the four-sphere partition function is dominated by
the classical term \eqref{Zcl}. After performing a change of variable, the
contribution from one-loop and instanton terms is negligible, and one recovers
the same Gaussian integral obtained for $\mathcal{N}=4$
\eqref{N=4-part-function} for each sector:
\begin{equation}\label{freeZamolodchikov}
	Z_{S^4}(\tau_i) \sim \prod_{a} Z^{\mathcal{N}=4}_{S^4}(\tau_a)\,,\qquad \text{as Im}\tau_a\to\infty\,.
\end{equation} 

We can now see that the behaviour we observed for $\mathcal{N}=4$ is very
generic in this limit. We can again construct infinite towers of composite
operators out of all possible fields of the theory. In the $\mathcal{N}=2$
case, we now have as many directions as there are gauge couplings---or
equivalently chiral multiplets of R-charge two---setting the dimension of the
conformal manifold. The relevant operators will be those made out of
combinations of $\ell$ appropriately-symmetrised derivatives and $n$ scalars
coming from the vector multiplets.\footnote{Note that these operators must be
	\emph{bona fide} conformal operators, i.e. eigen-operators of the
	dilatation generator, and there will in general be mixing between
operators with the same quantum numbers. This point is irrelevant to our
analysis, as we are only interested in the qualitative behaviour near the
infinite-distance point.} Their conformal dimensions is then
\begin{equation}
	\Delta_{\mathcal{O}_{n,\ell}} = n + \ell + \gamma(\tau_1,\dots,\tau_{\text{dim}\mathcal{M}}) \, .
\end{equation}
A particularity of operators constructed out of scalars coming from vector
multiplets is that any interaction term involving them will either come from
gauged kinetic terms or from F-terms, and therefore always involve powers of the
coupling. In the limit where $\text{Im}\tau_a\to\infty$, the anomalous
dimensions will be proportional to the gauge couplings, and using the form of
the Zamolodchikov metric \eqref{freeZamolodchikov} one finds:
\begin{equation}\label{ConformalDimensionAtInfinitePoint}
	\Delta_{\mathcal{O}_{n,\ell}} \sim  n + \ell + \eta\, 
	e^{-\alpha_{\chi}\,\text{dist}_\chi(\tau_a)} \, .
\end{equation}
Note that, while for $\mathcal{N}=4$ super-Yang--Mills the case-dependent
coefficient $\eta$ was a pure number, it now can depend on the other couplings that
are not taken to the free limit.

We therefore obtain the same qualitative behaviour observed for
$\mathcal{N}=4$: in the bulk there is an infinite tower of scalar fields of
mass $(m_nL)^2\sim n^2$, but more importantly, and as required by
Maldacena--Zhiboedov theorem \cite{Maldacena:2011jn}, there is also an infinite
number of higher-spin currents of the form \eqref{higherSpinCurrents} that
become conserved. The latter class of operators are again mapped to the bulk as
an infinite tower of higher-spin modes becoming exponentially light with the
distance as in \eqref{higher-spin-fields}, as required by the SDC. Similarly,
the density of the tower being linear with the spin, we can again expect
that these limits do not correspond to a partial decompactification in the
string theory description.

Since the tower of states satisfies the SDC, moving away from the ``interior''
of the moduli space towards an infinite-distance point, the mass of the
higher-spin states will eventually fall below the constant AdS radius.
Similarly to what happens in the case of the moduli space of
$\text{AdS}_5\times S^5$ discussed in the previous section, the supergravity
regime will again break down, and the appropriate description will be that of a
weakly-coupled CFT. As will be discussed shortly, superconformal representation
theory severely restricts the possible infinite-distance points of the
conformal manifold. This means that moduli spaces of consistent $\text{AdS}_5$
supergravity theories with sixteen supercharges does not contain any
infinite-distance points---at least of the type considered here---where an
effective description does not completely breaks down when taking into account
quantum corrections.

This breakdown is in spirit similar to what happens in flat space when trying
to reach the small-volume point of Calabi--Yau moduli spaces. As one tries to
approach it, one leaves geometric phase of the moduli space, and the usual
classical moduli are not appropriate quantum variables, leading to a quantum
obstruction. Such examples have been studied in the context of the SDC in
\cite{Baume:2019sry,Lee:2019wij}.

Conversely, an obvious difference with $\mathcal{N}=4$ is that many of the
operators will not go to their free value, even when they contain fields
charged under the gauge group that decouples. The anomalous dimension of such
composite operators will generically not be proportional to the associated
couplings and there can be mixing with fields of another sector, if for
instance they are in the bifundamental representation of groups whose coupling
does not go in the free limit.

Having found a tower of states compatible with the SDC, we can now inquire about
the order of magnitude of the exponential rate, $\alpha_{G}$. One can consider
several sectors decoupling at different paces, and this will be reflected in
the value of $\alpha_G$. Let us introduce a parameter, $t$, describing the
fastest gauge couplings satisfying $\text{Im}\tau_a\to \infty$. Those going to
the same limit, but slower, can be described similarly by introducing an
exponent, $p_a$:
\begin{equation} \label{eq:limits}
	\text{Im}(\tau_{a}) = t^{\,p_{a}}, \quad 0<p_{a}\leq 1 \,.
\end{equation}
Of course, $p_a=1$ only for the parameter---or family of parameters---going to
the free limit the fastest. We note that all these trajectories are
geodesics, as can be seen by using flat coordinates $\Phi_a \sim
\log(\text{Im}\tau_{a} )$ with respect to the Zamolodchikov metric derived from
\eqref{freeZamolodchikov} and checking that they are straight lines.

Using \eqref{freeZamolodchikov} and \eqref{eq: Zamo-partition-function} one can
estimate the distance in the Zamolodchikov metric in terms of this parameter,
and then translate it to a distance in the moduli space using
\eqref{eq:metric-dictionary}. One proceeds in the same fashion as for
$\mathcal{N}=4$ to obtain the usual logarithmic behaviour, and a decay rate,
\begin{equation}\label{eq:decay_rate_N=2}
	\alpha_{G} \sim \left(\frac{(LM_\text{Pl})^3}{\sum_{a}p_{a}^{2}\, \text{dim}(\mathcal{G}_{a})}\right)^{\frac{1}{2}} \,.
\end{equation}
When the theory admits a point in the moduli space where a supergravity
description in terms of Einstein gravity is available, we can estimate the
value of $LM_\text{Pl}$ by computing the trace-anomaly coefficients of the CFT,
$a, c$. Going through the usual holographic computation, one obtains that at
leading order in $N$, $(LM_\text{Pl})^3\sim a$. The coefficients further agree
up to linear corrections in $N$, $24(a-c)=n_v-n_h= \mathcal{O}(N)$, where
$n_h\,,n_v$ correspond to the number of $\mathcal{N}=2$ hyper- and vector
multiplets, respectively. For large-enough values of $N$, the standard formulas
therefore yield:
\begin{equation}
	a = \frac{5n_v+n_h}{24} \sim \frac{n_v}{4} = \frac{1}{4}\text{dim}(\mathcal{G})\,.
\end{equation}
Further gravitational corrections in the bulk will modify the value of the trace-anomaly
coefficients, but those will always be subleading in $N$ and will not modify
the overall scaling. For our purpose, we can therefore use them to estimate the
scaling of $(LM_\text{Pl})^3$ in terms of the dimension of the total gauge
group:
\begin{equation}
	\alpha_{G} \sim \left(\frac{\text{dim}(\mathcal{G})}{\sum_{a}p_{a}^{2}\,\text{dim}(\mathcal{G}_{a})}\right)^{\frac{1}{2}} \, .
\end{equation}
We see that the denominator is bounded between $\text{dim}(\mathcal{G}_{dec.})$, the
dimension of the gauge subgroup decoupling the fastest which by assumption has
$p_a=1$, and the dimension of the total gauge group. For any free limit and
large-enough groups, we find the bounds: 
\begin{equation}\label{DecayBound}
	\mathcal{O}(1) \lesssim\, \alpha _{G}\, \lesssim \left(\frac{\text{dim}(\mathcal{G})}{\text{dim}(\mathcal{G}_\text{dec.})}\right)^{\frac{1}{2}} \, .
\end{equation}

This means that the exponential rate is always of order one in Planck units, or
larger. It is thus very easy to engineer limits with large $\alpha_{G}$. For
example a theory with gauge group $SU(N)^{K}$ in the limit where a single
$SU(N)$ becomes free leads to:
\begin{equation}
	\alpha_{G} \sim \sqrt{K} \, .
\end{equation} 

Note that while we focussed on $\mathcal{N}=2$ four-dimensional theories where
the relations between the trace-anomaly coefficients and the gauge group data
are simple, estimating $LM_\text{Pl}$ in terms of group theoretical data of the
gauge theory can be adapted \emph{mutatis mutandis} to studies in other
dimensions, trading $a,c$ for the appropriate quantities. We therefore expect
similar bounds in more general cases whenever a sector of the CFT decouples.

\subsection{Beyond Free Points}\label{sec:BeyondFreePoints}

If the Swampland Distance Conjecture is true, we expect infinite-distance points
to be associated with an infinite towers of massless states in the bulk. As we
have seen, those associated to a weak-gauge-coupling limit on the boundary CFT
will have an infinite number of higher-spin conserved currents, as required by
the Maldacena--Zhiboedov theorem. One may then ask what type of behaviour one
can expect beyond those where a sector becomes free, if any.

For instance, one can consider a limit in which a tower of scalars become
massless in the bulk and whether it is at infinite distance. Via the dictionary
\eqref{dicScalar} such a tower can only appear if there are points with
additional marginal operators in the boundary, which by $\mathcal{N}=2$
superconformal representation theory only exists when there is a gauge symmetry
enhancements of the CFT. We can always move slightly away from the conformal
manifold onto the Coulomb branch by giving a vacuum expectation value to scalar
fields inside the vector multiplet. As the dimension of the Coulomb branch is
an invariant of the theory, the total rank of the gauge group is fixed and an
infinite tower of scalars is not possible. Beyond $\mathcal{N}=2$, we are not
aware of any CFT exhibiting loci in the conformal manifold where an infinite
number of new marginal deformations appear. These towers would be ideal
candidates for Kaluza--Klein towers in the bulk and their apparent absence
again provides support to the expectation that the ADC and SDC directions in
moduli space are separate limits. 

Another possibility is an enhancement of the flavour group of the CFT. This
requires a would-be flavour current to be part of a long multiplet that becomes
short. As shown in \cite{Beem:2013sza,Beem:2014zpa}, an analysis of the
recombination rules of long multiplets at threshold reveal the only such
possibility to be a superconformal multiplet of type
$\widehat{\mathcal{C}}_{0,(\frac{1}{2},\frac{1}{2})}$. This multiplet contains
a higher-spin conserved current, implying that there is again a sector of the
SCFT that will decouple. It in turn means that the associated gauge
enhancements in the bulk are at infinite distance.

There are also strongly-coupled points in the conformal manifold that
are at infinite distance. These points are often free points in disguise, as
there exists a duality transformation to a frame where there is a
weakly-coupled sector. Such examples are plentiful in class S constructions,
and we will consider specific cases in the next section.

While we are not able to show that there are no infinite-distance point that do
not correspond to a decoupling limit of a $\mathcal{N}=2$ SCFT, we are not aware
of such a case. Using localisation techniques, it is in principle feasible to
compute the Zamolodchikov metric in a non-perturbative regime by taking into
account all loop and instanton corrections in \eqref{PartitionFunctionContribution}.

Finally, there cannot be compact smooth conformal manifolds with
$\mathcal{N}=2$ supersymmetry \cite{Gomis:2015yaa}, thereby excluding cases
that do not admit any free limit at all. There is furthermore a conjecture
stating that any $n$-dimensional $\mathcal{N}=2$ conformal manifold can be
obtained by gauging $n$ simple factors of the flavour symmetry associated to
SCFTs with no marginal deformations \cite{Beem:2014zpa}. In that sense, all the
infinite-distance points studied in this work correspond to reversing
(partially or completely) the process by returning to a flavour symmetry.

We close this section by noting that the results we have obtained carry to cases
with lower dimensions and supersymmetry. Whenever a sector of the CFT becomes
free there will always be a tower of massless higher-spin fields in the bulk.
However, this does not mean that sending any marginal coupling to zero will
involve a tower of the form \eqref{higherSpinCurrents}. Indeed let us imagine an
$\mathcal{N}=1$ SCFT depending on two marginal parameters. As marginal operators
need not be gauge deformations in that case, sending one of the parameters to
zero does not imply that the anomalous dimensions of would-be conserved currents also
vanish. It might still depend non-trivially on the other parameter, depending on
the structure of the CFT, and the decoupled point could be at finite distance.
While there is a possibility that it may be at infinite distance and an SDC
tower still exists, it requires a further analysis of $\mathcal{N}=1$
superconformal representations, which we leave for future works.

\section{Orbifolds and $\mathcal{N}=2$ Necklace Quivers}
\label{sec:orbifolds}

In Section \ref{sec:CFTs} we have reviewed the machinery of four-dimensional
$\mathcal{N}=2$ SCFTs to learn about the possible behaviour of SDC towers in
$AdS_5$ vacua with sixteen supercharges. To understand the mechanisms
responsible for the associated infinite-distance points in the bulk, as well as
exploring points that are a priori not free, we now turn to an explicit
construction in string theory, namely the family of $\text{AdS}_5$ vacua
obtained by type IIB compactification on an orbifold of the form $S^5/\Gamma$.
In order to conserve sixteen supercharges we are forced to consider the
orbifold action to be an ADE-type discrete subgroup, $\Gamma\subset SU(2)$.
For simplicity we focus on the A-type series, that is, on backgrounds of the
form $AdS_5 \times S^5/\mathbb{Z}_K$.\footnote{Seen as a constant-radius
	hypersurface on $\mathbb{C}^3$, the precise action of $\mathbb{Z}_k$ on
	the $S^5$ is given by $\left(z_1,z_2,z_3\right) \rightarrow \left(
e^{2\pi i/K}z_1, e^{-2\pi i/K}z_2, z_3 \right)$.} Other cases can be
generalised straightforwardly.

\begin{figure}[t]
	\centering
	\resizebox{.3\textwidth}{!}{\Huge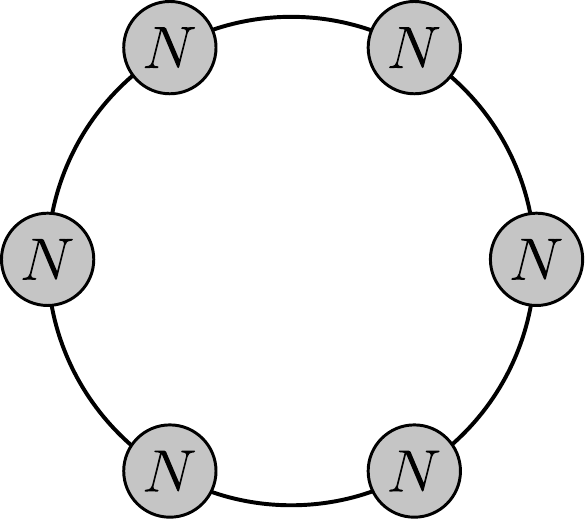}
	\qquad
	\resizebox{.4\textwidth}{!}{\Huge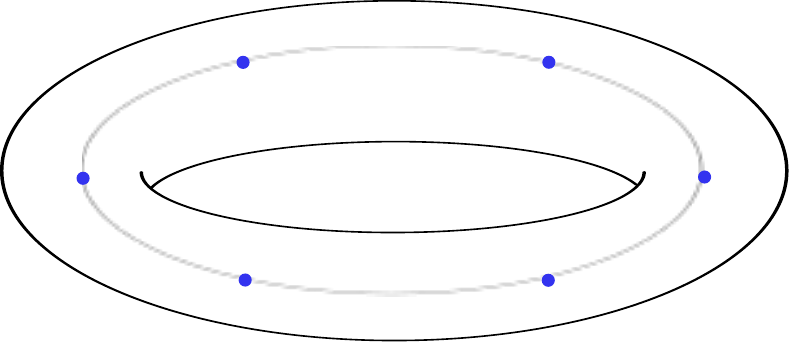}
	\caption{Left: quiver representation for the $\hat{A}_{K-1}$ necklace
	theory, with each node corresponding to a $\mathcal{N}=2$ vector
	multiplet and each line a hypermultiplet in the bifundamental of the adjacent
	groups. Right: the torus with $K$ minimal punctures, $T^2_K$, of the
	associated class S construction.}
	\label{fig:quiver}
\end{figure}

The dual four-dimensional $\mathcal{N}=2$ SCFT are the well-known necklace
quiver theories with gauge group $\mathcal{G}=SU(N)^K$
\cite{Lawrence:1998ja,Kachru:1998ys}, which are represented by the affine
Dynkin diagram $\hat{A}_{K-1}$, as depicted in figure \ref{fig:quiver}. In
addition to vector multiplets associated to each gauge factor, there are
also hypermultiplets transforming in bifundamental representations of
each pair of adjacent gauge factors. At large $N$, their Zamolodchikov metric
was studied in \cite{Pini:2017ouj}.

Necklace theories can be obtained by projecting out modes of $\mathcal{N}=4$
super-Yang--Mills theory with gauge group $SU(KN)$ and it is natural that the
complexified gauge coupling of each gauge sector, $\tau_i$, is related to that
of the parent theory, $\tau_0$ \cite{Lawrence:1998ja}:
\begin{equation} \label{orbifold-point-CFT}
	\tau_{i} = \frac{\tau_0}{K}\,,\qquad \tau = \sum_{i=1}^K \tau_i \,.
\end{equation}

Recalling that $\tau_0=\tau$ is the holographic dual of the axio-dilaton, we
see that it is controlling the overall complex gauge coupling in this setup. As
each gauge coupling, $\tau_i$, is again associated to a marginal deformation,
we expect $K-1$ additional complex moduli in the bulk. To separate these from
the axio-dilaton, we choose the following basis of marginal couplings:
\begin{equation} \label{eq:basis-CFT}
	\tau_i\in\{\tau_0, \tau_{a}\}\,,\qquad a = 1, \ldots, K-1.
\end{equation}
Since the axio-dilaton is the only moduli of the unorbifolded theory, the other
moduli should come from the twisted sector. Indeed, $\mathbb{Z}_K \subset
SU(2)$ does not act freely on the five-sphere, but leaves a circle
$S^{1}\subset S^{5}$ of singular fixed points after orbifolding. This leads to
a twist sector related to the $K-1$ blow-up 2-cycles resolving these
singularities. One finds five scalars for each 2-cycle: two axions and three
of geometric origin. It turns out that the latter are stabilised by the
potential, and consequently are not part of the moduli space. Thus, as proposed in
\cite{Lawrence:1998ja}, the $K-1$ complex moduli we are looking for are the
axions coming from the periods of the RR and NSNS 2-forms on the 2-cycles,
$b_{a}$ and $c_{a}$ respectively. We will assume them to be normalised to have
period one. All in all, one finds the following dictionary between the moduli
and the marginal couplings:
\begin{equation} \label{eq:complex-couplings}
	\tau_{a} = c_{a} + \tau_0\,b_{a}\,,
\end{equation}
or in terms of Yang--Mills gauge couplings and theta-angles,
\begin{equation} \label{eq:coupling-angles}
    \frac{\theta_{a}}{2\pi}= c_{a} + b_{a} C_{0}\,,\qquad
    \frac{4\pi}{g_{a}^{2}} = \frac{b_{a}}{g_{s}}\,. 
\end{equation}
Additional details on the duality between axions in the twisted sector and
marginal couplings can be found in
\cite{Katz:1997eq,Witten:1997sc,Gukov:1998kk}.

Notice that tuning the vevs of these twist-sector moduli corresponds to a
deformation of the theory away from the one obtained by orbifolding
$\text{AdS}_{5}\times \text{S}^{5}$. Indeed, \eqref{orbifold-point-CFT} is only
recovered when the vev of these axions correspond to the orbifold point
\cite{Aspinwall:1994ev},
\begin{equation} \label{eq:orbifold-point}
	c_{a} = 0, \qquad b_{a} = \frac{1}{K}.
\end{equation}
There are various ways to reach an infinite-distance point by tuning the
different moduli, which as we will see will lead us away from the orbifold point
to regions where the supergravity description breaks down. The situation is
similar to that of four-dimensional $\mathcal{N}=2$ theories in flat space, as
we can tune one or more moduli at the same time, leading to different
behaviours. We now analyse various limits and try to identify the associated
infinite towers of states, both in the bulk and the CFT.

\subsection{Overall Free Limit}\label{sec:overall-free}

From the definition of the moduli $\tau_i$ it is straightforward to see that
the simplest limits are the ones for which all the gauge sectors are becoming
free at the same rate. In the bulk, this is controlled by the limit
\begin{equation}
	\text{Im}(\tau) \rightarrow \infty\, ,
\end{equation}
while $b_a$, $c_a$ and $C_0$ are fixed. From the expression of the
Zamolodchikov metric \eqref{freeZamolodchikov}, we see that this point is indeed
at infinite distance, and we are in a situation similar to that of section
\ref{sec:warm-up}, where the fundamental Type IIB string becomes tensionless and
weakly coupled. In this particular case, we can also make use of type IIB
$SL(2,\mathbb{Z})$ duality to argue that the same behaviour occurs when reaching
the overall strong coupling limit, $\text{Im}\tau\to0$ with all the other
moduli constant.

There is again an infinite tower of higher-spin fields, dual to generalised
currents, $J_{\mu_1\dots\mu_\ell}$, becoming exponentially massless with the
distance in Planck units:
\begin{equation}
	\frac{M_{\ell}}{M_\text{Pl}}\sim e^{-\alpha_{G}\,\text{dist}_{G}(\tau)} \, .
\end{equation}
These massless higher-spin currents have an obvious interpretation in the bulk:
they originate from higher-spin excitations of tensionless fundamental
strings.  Moreover, being an overall free limit in the CFT with all gauge
sectors decoupling at the same rate, we know from the discussion in section
\ref{sec:The-Zamo-Metric} that $\alpha_{G}$ is of order one. 

Relaxing the condition $b_a=\text{const}$, it is possible to also explore
limits in which different $SU(N)$ sectors decouple at different rates. For this
more general case we know from \eqref{DecayBound} that $\alpha_{G}$ will be
bounded between an order one number and $\sqrt{K}$, and can as a result be
parametrically large.

As in $\mathcal{N}=4$ SYM, but contrary to Minkowski backgrounds, only
a small number of scalar fields become massless, while using
\eqref{ConformalDimensionAtInfinitePoint} and the AdS/CFT dictionary, all other
scalars have masses that are regularly spaced near the infinite-distance point:
\begin{equation}
	\frac{M_n}{M_\text{Pl}}\sim n^2\,.
\end{equation}

This case is however quite special, as everything is controlled by the value of
the axio-dilaton and all fundamental fields of the CFT are free in that limit.
We can reach a much richer network of infinite-distance point in moduli
space by demanding that only a strict subset of the moduli change over the
path, which in the CFT corresponds to taking only some of the gauge couplings
to be free.

\subsection{Strong-coupling Points and Dualities}
\label{sec:strong-limits}

Demanding the axio-dilaton to remain constant, tuning some---or all---of the
$K-1$ remaining moduli to zero, we can reach a variety of new points in the
conformal manifold:
\begin{equation}\label{eq:strongCouplingLimitOrbifold}
	\tau = \text{const}\,, \qquad
	b_a, c_a \rightarrow 0\,.
\end{equation}
One could naively expect this path to lead to a finite-distance point, as we
are moving in axionic directions. However, moving away from the orbifold point
we cannot trust the usual supergravity description, but we can use the dual CFT
to nonetheless learn about what happens in its neighbourhood.

Using the dictionary \eqref{eq:coupling-angles}, sending the moduli to zero
corresponds to keeping the overall complex coupling fixed while taking the
remaining $K-1$ gauge sectors to strong coupling:
\begin{equation}
	\tau_0=\text{const}\,,\qquad
	g_a \to \infty\,, \qquad 
	\theta_a \to 0\,.
\end{equation}
Note that this limit also demands the $\theta$-angles to fall to zero. This
setup was studied in details in \cite{Aharony:2015zea} using Gaiotto's class S
construction. In that framework, the necklace quivers are realised as a
compactification of the six-dimensional $\mathcal{N}=(2,0)$ SCFT of type
$A_{N-1}$ on a torus with $K$ minimal regular punctures, $T^2_K$, depicted in
figure \ref{fig:quiver}.

From this point of view, a trajectory in the conformal manifold in which the
axio-dilaton is kept fixed corresponds to changing the relative position of the
punctures, and the limit above brings them all together, possibly at
different rates. A key point of the class S framework is that as one brings two
punctures together, the torus develops a throat and there exist an S-duality
frame in which a sector of the theory becomes a weakly-coupled gauge theory.
Furthermore, this process is local, and does not depend on what happens in the
rest of the surface \cite{Chacaltana:2010ks}.

In the case at hand, one obtains a sector consisting of a strongly-interacting
SCFT, where the $\mathcal{N}=(2,0)$ theory is compactified on a torus with a
single puncture, connected to a weakly-coupled gauge theory with gauge group
\cite{Aharony:2015zea}:
\begin{equation}
	\mathcal{G}_\text{dec.}\subset\widehat{\mathcal{G}} = SU(2)\times SU(3)\times \dots\times SU(K)\,.
\end{equation}
We are therefore once again in the situation described around equation
\eqref{ConformalDimensionAtInfinitePoint}: the conformal dimensions of
operators built out of fields in the weakly-coupled vector multiplets have an
anomalous dimension that is exponentially suppressed with the distance around
the point \eqref{eq:strongCouplingLimitOrbifold} in this particular duality
frame, which is at infinite distance. The Maldacena--Zhiboedov theorem again
requires the presence of an infinite tower of higher-spin conserved currents.

In the bulk, we also have an infinite tower of higher-spin operators that
become massless exponentially fast in Planck units, $M_\ell/M_\text{Pl}\sim
e^{-\alpha_G\,\text{dist}_G}$. The decay rate can then be estimated using the
bounds \eqref{DecayBound} in terms of the group theory data of the dual.

Let us comment on the stringy origin of these states. As $b_a$ and $c_a$ become
small, so does the tension of D3-branes wrapped on the blow-up cycles:
\begin{equation}
	T_\text{D3}\sim \int_{\Sigma_a}\left|C_2+\tau B_2\right| \xrightarrow{~b_a,c_a\to0~}0\,.
\end{equation}
One might be tempted to conclude that the massless higher-spin fields in the
bulk come from such tensionless strings, particularly in the context of the
Emergent String Conjecture \cite{Lee:2018urn}. However, little is known about the spectrum of
these strings in AdS---in particular, in flat space they are non-critical and
do not give rise to an infinite number of massless states---and we are
therefore unable to make such a claim. The origin of the tower of states
remains elusive in these limits. 

Intriguingly, it was proposed in
\cite{Aharony:2015zea} that there might be a dual description where there is no
tower of higher-spin modes in the bulk. There, part of the SCFT is
associated with a four-dimensional strongly-interacting gauge theory living on
the boundary of AdS${}_5$, which is then coupled to the rest of the bulk, i.e.
to the rest of the type IIB spectrum. At the point where the boundary SCFT
becomes free, this gauge theory becomes weakly-coupled. Therefore the
higher-spin conserved currents in the boundary are not mapped to massless
higher-spin modes in the bulk, but to the higher-spin conserved currents of this
four-dimensional gauge theory. This possibility however involves choosing
non-standard boundary conditions, and the usual AdS/CFT dictionary does not
apply. We leave an exploration of such a description and its relation to the SDC
for future works.

Classically, the limits we have been considering would be at finite distance and
we expect the infinite distance to be driven by quantum
corrections.\footnote{This can be seen by considering the moduli space as a
truncation of the theory obtained by placing the orbifold in flat space. For the
latter, the moduli space is classically exact and our limits are known to be at
finite distance \cite{Aharony:2015zea}.} We note that, similarly to what happens
in the case of $\text{AdS}_5\times S^5$, as we move away from the orbifold point
by tuning the axions, we leave the phase of the moduli space where the
supergravity regime is valid, and enter a phase where the CFT description is
more appropriate.

The behaviours described above generalise to a wide zoo of class S examples.
Given a $\mathcal{N}=(2,0)$ six-dimensional SCFT of ADE type, one can reach a
four-dimensional $\mathcal{N}=2$ SCFT by compactifying on a punctured compact
Riemann surface \cite{Gaiotto:2009we}, see \cite{Tachikawa:2013kta} for a
review. The surfaces can then be constructed as 3-punctured spheres glued by
tubes, called ``tinkertoys'' \cite{Chacaltana:2010ks, Chacaltana:2012ch}. For
SCFTs of type $A_N$, the allowed collisions of punctures leading to a
weakly-coupled gauge sector have been studied in \cite{Genish:2017cdp}. In some
cases, one can construct a weakly-curved holographic gravity dual from M-theory
on a background of the form $\text{AdS}_5\times X_6$. As for
$\text{AdS}_5\times S^5$ and its orbifolds, the infinite-distance points will be on the
boundary of the moduli space where there supergravity regime has broken down
because the tower of state has reached a scale smaller than the AdS radius.

\section{Conclusions}\label{sec:conclusions} 

By studying the behaviour of states near a family of infinite-distance points
in the moduli space of AdS vacua, we have taken a first step towards a possible
extension of the Swampland Distance Conjecture to curved backgrounds. To that
end, we have used the power of the dual four-dimensional $\mathcal{N}=2$
superconformal symmetry, which allows one to reduce large classes of
infinite-distance points to a case where a subsector of the SCFT becomes free.
While we are unable to claim that all infinite-distance points correspond to
free limits, we are not aware of possible counterexamples. In the bulk, there
is then always an infinite number of higher-spin modes becoming exponentially
massless playing the r\^ole of the SDC tower for AdS moduli spaces. 

In flat space, the tower of light states indicates that the effective field
theory description is no longer valid, and the SDC is parametrising how this
breakdown occurs. By contrast, we find that in the bulk this always happens
before reaching the infinite-distance point. For instance, this limit for
$\text{AdS}_{5}\times \text{S}^{5}$ vacua is located in the highly-curved
regime. On general grounds, the SDC predicts that a tower of states will
eventually fall below the AdS scale and an effective description would need a
lower cut-off. Should this not be an appropriate description, it would mean
that the landscape of AdS vacua in quantum gravity cannot admit an effective
field theory description when getting close to infinite-distance points in
moduli space, thereby strongly constraining the possible theories which can be
coupled to quantum gravity. Note that how close to the infinite-distance point
one can go with an effective theory depends on the AdS radius. In particular,
it would be interesting to relate the lack of effective description to the
species scale, as is done in flat space \cite{Heidenreich:2016aqi}. In this
context, the species scale controls the effective gravity coupling, and thus the
size of a typical quantum fluctuation around the background metric. If this
logic applies to AdS, when the species scale becomes smaller than the AdS scale,
they are large compared to the background, making  a geometric description
inconsistent.

For theories described by Einstein gravity at a point of the moduli space we
have also been able to find bounds for the exponential decay constant. It must
always be at least of order one in Planck units and is bounded from above by
the ratio between the dimensions of the total gauge group and the decoupled
sector. It is therefore possible to obtain a parametrically-large decay
constant by engineering a small sector decoupling from a large gauge group.

We have applied this analysis to orbifolds of $S^5$, with the associated CFT
being described by necklace quivers. When all gauge nodes are decoupled, one
finds a tensionless fundamental string in the bulk. Using the class-S
description of that SCFT we were further able to relate the behaviour of the
SDC for individual strong-coupling points to that of free limits via S-duality
and found no other infinite-distance points. However, the stringy
interpretation of the tower of states is less obvious in these cases: at these
points, D3-branes wrapped on blow-up two-cycles become tensionless, but their
flat-space avatars are non-critical strings and at finite distance in moduli
space. Further, the effective description has broken down well before reaching
that point. A relation with the Emergent String Conjecture \cite{Lee:2018urn}
is therefore not conclusive, and calls for further analysis.

Many of the arguments we discussed here generalise to more arbitrary cases,
with and without supersymmetry, that admit a CFT dual. The Maldacena--Zhiboedov
theorem does not require supersymmetry and there will always be an infinite
tower of higher-spin states in a limit leading to subsector of the CFT becoming
free. If the marginal deformation is identified with a gauge coupling along the
conformal manifold, it will be at infinite distance by looking at the
Zamolodchikov metric and there will be exponentially light states accompanying
it. However, the structure of conformal manifolds greatly depends on the
spacetime dimension and number of supercharges. For instance, there are no
supersymmetry-preserving marginal deformations in six dimensions, which in the
bulk translates to all moduli being stabilised \cite{Apruzzi:2013yva}. In lower
dimensions however, there can be marginal deformations that go beyond changing
gauge coupling constants. One would therefore expect to have a far richer
network of infinite-distance points in these cases. It might  be very
interesting to look for similar structures as the ones used in the context of
Hodge theory, see e.g.
\cite{Grimm:2018ohb,Grimm:2018cpv,Grimm:2019ixq,Grimm:2020cda}.

Furthermore, there also exists conformal manifolds which are compact, see e.g.
\cite{Buican:2014sfa}. While their holographic duals are not well understood,
it would be interesting to see if the requirements needed to have a compact
manifold can be related to swampland constraints in the bulk.

Unlike the Weak Gravity Conjecture, where black hole physics plays an important
r\^ole, the current understanding of the SDC comes principally from string
theory. Using the AdS/CFT correspondence therefore opens new avenues to explore
this part of the Swampland programme. For instance, unitarity constraints and
other features of superconformal symmetry might shed new light on the origin of
the various conjectures and how they are related.

\subsection*{Acknowledgements}

We thank J. Barb\'on, C. Lawrie, M. Montero, J. Quirant, A. Uranga, I.
Valenzuela, and M. Wiesner for helpful discussions and comments on the
manuscript. The work of F.B. is supported by the Spanish Research Agency
(Agencia Estatal de Investigaci\'on) through the grant IFT Centro de Excelencia
Severo Ochoa SEV-2016-0597, by the grant PGC2018-095976-B-C21 from
MCIU/AEI/FEDER, UE, and the Swiss National Science Foundation (SNSF) grant
number P400P2\_194341. J.C. is supported by the FPU grant no. FPU17/04181 from
Spanish Ministry of Education.

\appendix

\section{Superconformal Representations of $SU(2,2|2)$}\label{app:representations}

We summarise in this appendix the basic notions of superconformal
representations useful in this work. The $\mathcal{N}=2$ superconformal group
is $SU(2,2|2)$, its bosonic subgroup being constituted of both the conformal
and R-symmetry groups, $SO(2,4)\times SU(2)_R\times U(1)_R$. The algebra
contains the usual conformal generators, namely those of translations,
rotations, and so-called special conformal transformations, $M_{\mu\nu},P_\mu,
K_\mu$, as well as the dilatation operator, $\mathcal{D}$. With the R-symmetry
generators, it is supplemented by the super- and superconformal charges:
\begin{equation}
	Q_{\alpha A}\,,\bar{Q}_{\dot{\alpha}A}\,,\quad
	S^{\alpha A}\,,\bar{S}^{\dot{\alpha}A}\,\qquad
	\alpha\,,\dot{\alpha} = 1,2\,,\quad A=1,\dots,4\,.
\end{equation}
One can then label any operator of the theory by the following quantum numbers:
\begin{equation}
	[\Delta; j,\bar{\jmath}; R; r]\,.
\end{equation}
In addition to the usual conformal dimension, $\Delta$, which is the charge of
the operator under dilatations, and the Lorentz Dynkin indices,
$(j,\bar{\jmath})$, of
$\mathfrak{so}(1,3)=\mathfrak{su}(2)\oplus\mathfrak{su}(2)$, we also define the
R-charges, $(R,r)$, of $SU(2)_R\times U(1)_R$.

The spectrum of the theory organises itself into superconformal multiplets
of $SU(2,2|2)$ whose highest weights are called superconformal primaries. A
superconformal primary, $\mathcal{O}$, is then by definition annihilated by
special conformal transformation generators---as for all usual conformal
primaries---and superconformal charges:
\begin{equation}
	[K_\mu,\mathcal{O}]=0\,,\qquad
	\left[ S^{\alpha A},\mathcal{O}  \right\}=0=\left[ 
	\bar{S}^{\dot{\alpha}A},\mathcal{O}  \right\}\,,\qquad
	\alpha,\dot{\alpha}=1,2\,,\quad A=1,\dots,4\,
\end{equation}
where the commutator or anti-commutator is used depending on whether
$\mathcal{O}$ is fermionic or bosonic. Given a superconformal primary, the rest
of the multiplet (called descendants), is then generated by successive
applications of the translation operator, $P_\mu$, and the regular
supercharges, $Q_{\alpha A}\,,\bar{Q}_{\dot{\alpha} A}$. Note that due to their
fermionic nature, the number of states generated by the supercharges is finite.

Moreover using the conformal algebra it possible to show that applying the
shift of conformal dimension of descendants from its primary is
(half-)quantised:
\begin{equation}
	\Delta_{[Q,\mathcal{O}]} = \Delta_{\mathcal{O}}+\frac{1}{2}\,,\qquad
	\Delta_{[P,\mathcal{O}]}=\Delta_\mathcal{O}+1\,.
\end{equation}
Schematically for a bosonic superconformal primary, the descendants and their
dimensions are found to be:
\begin{equation}
	\begin{aligned}[t]
	\mathcal{O}\,,\quad&
	[Q_{\alpha A},\mathcal{O}]\,, \quad&
	[\bar{Q}_{\dot{\alpha} A},\mathcal{O}]\,, \quad&
	\dots&
	[P_\mu, \mathcal{O}]\,,\quad&
	[P_{\nu},[P_\mu, \mathcal{O}]]\,,\quad&
	\dots\\
	\Delta\,,\quad&
	\Delta+\frac{1}{2}\,,&
	\Delta+\frac{1}{2}\,,~\quad&
	\dots\quad&
	\Delta+1\,,\quad&
	\quad\Delta+2\,,\quad&
	\dots
	\end{aligned}
\end{equation}
Conversely, for superconformal charges and special conformal transformations,
the sign of the shift is reversed.

A complete analysis of the representations then separates superconformal
multiplets into two classes: long and short multiplets. The latter corresponds
to cases where the superconformal primary is annihilated by a combination of
the supercharges, in which case its dimension is set by the rest of the quantum
numbers. In this work, we are mainly interested in a class of multiplet whose
superconformal primary is annihilated by half of the supercharges, such as the
chiral multiplets, $\mathcal{E}_r$. Long multiplets are unconstrained and their
conformal dimensions are only bounded from below by unitarity.

Additional details and a complete classification of unitary superconformal
multiplets can be found in e.g. \cite{Dolan:2002zh,Cordova:2016emh}.

\newpage
\bibliography{references}{}
\bibliographystyle{JHEP} 

%
\end{document}

%% file: quiverN2.pdf_tex
\begingroup%
  \makeatletter%
  \providecommand\color[2][]{%
    \errmessage{(Inkscape) Color is used for the text in Inkscape, but the package 'color.sty' is not loaded}%
    \renewcommand\color[2][]{}%
  }%
  \providecommand\transparent[1]{%
    \errmessage{(Inkscape) Transparency is used (non-zero) for the text in Inkscape, but the package 'transparent.sty' is not loaded}%
    \renewcommand\transparent[1]{}%
  }%
  \providecommand\rotatebox[2]{#2}%
  \newcommand*\fsize{\dimexpr\f@size pt\relax}%
  \newcommand*\lineheight[1]{\fontsize{\fsize}{#1\fsize}\selectfont}%
  \ifx\svgwidth\undefined%
    \setlength{\unitlength}{168.24446086bp}%
    \ifx\svgscale\undefined%
      \relax%
    \else%
      \setlength{\unitlength}{\unitlength * \real{\svgscale}}%
    \fi%
  \else%
    \setlength{\unitlength}{\svgwidth}%
  \fi%
  \global\let\svgwidth\undefined%
  \global\let\svgscale\undefined%
  \makeatother%
  \begin{picture}(1,0.88787025)%
    \lineheight{1}%
    \setlength\tabcolsep{0pt}%
    \put(0,0){\includegraphics[width=\unitlength,page=1]{quiverN2.pdf}}%
  \end{picture}%
\endgroup%

%% file: torus.pdf_tex
\begingroup%
  \makeatletter%
  \providecommand\color[2][]{%
    \errmessage{(Inkscape) Color is used for the text in Inkscape, but the package 'color.sty' is not loaded}%
    \renewcommand\color[2][]{}%
  }%
  \providecommand\transparent[1]{%
    \errmessage{(Inkscape) Transparency is used (non-zero) for the text in Inkscape, but the package 'transparent.sty' is not loaded}%
    \renewcommand\transparent[1]{}%
  }%
  \providecommand\rotatebox[2]{#2}%
  \newcommand*\fsize{\dimexpr\f@size pt\relax}%
  \newcommand*\lineheight[1]{\fontsize{\fsize}{#1\fsize}\selectfont}%
  \ifx\svgwidth\undefined%
    \setlength{\unitlength}{227.10118408bp}%
    \ifx\svgscale\undefined%
      \relax%
    \else%
      \setlength{\unitlength}{\unitlength * \real{\svgscale}}%
    \fi%
  \else%
    \setlength{\unitlength}{\svgwidth}%
  \fi%
  \global\let\svgwidth\undefined%
  \global\let\svgscale\undefined%
  \makeatother%
  \begin{picture}(1,0.43287411)%
    \lineheight{1}%
    \setlength\tabcolsep{0pt}%
    \put(0,0){\includegraphics[width=\unitlength,page=1]{torus.pdf}}%
  \end{picture}%
\endgroup%